\documentclass
[superscriptaddress,secnumarabic,amssymb,amsmath,nobibnotes,aps,prd,showkeys,showpacs,nofootinbib,twocolumn,notitlepage,10pt]{revtex4}%
\usepackage{setspace}
\usepackage{color}
\usepackage{amsmath}
\usepackage{amsfonts}
\usepackage{verbatim}
\usepackage{amssymb}
\usepackage{graphicx,bm}
\usepackage{amsmath}
\usepackage{amssymb}
\usepackage{graphicx}%
\providecommand{\U}[1]{\protect\rule{.1in}{.1in}}
\newcommand{\be}{\begin{equation}}
\newcommand{\ee}{\end{equation}}

\newcommand{\mincir}{\raise
-3.truept\hbox{\rlap{\hbox{$\sim$}}\raise4.truept\hbox{$<$}\ }}
\newcommand{\magcir}{\raise
-3.truept\hbox{\rlap{\hbox{$\sim$}}\raise4.truept\hbox{$>$}\ }}

\ifx\pdfoutput\relax\let\pdfoutput=\undefined\fi
\newcount\msipdfoutput
\ifx\pdfoutput\undefined\else
\ifcase\pdfoutput\else
\ifx\paperwidth\undefined\else
\ifdim\paperheight=0pt\relax\else\pdfpageheight\paperheight\fi
\ifdim\paperwidth=0pt\relax\else\pdfpagewidth\paperwidth\fi
\fi\fi\fi
\begin{document}
\title{Testing Non-Coincident $f(Q)$-gravity with DESI DR2 BAO and GRBs}
\author{Andronikos Paliathanasis}
\email{anpaliat@phys.uoa.gr}
\affiliation{Department of Mathematics, Faculty of Applied Sciences, Durban University of
Technology, Durban 4000, South Africa}
\affiliation{Centre for Space Research, North-West University, Potchefstroom 2520, South Africa}
\affiliation{Departamento de Matem\`{a}ticas, Universidad Cat\`{o}lica del Norte, Avda.
Angamos 0610, Casilla 1280 Antofagasta, Chile}
\affiliation{National Institute for Theoretical and Computational Sciences (NITheCS), South Africa}

\begin{abstract}
We consider the $f\left(  Q\right)  $-theory for the description of dark
energy with a non-trivial connection defined in the non-coincident gauge. The
resulting field equations form a two-scalar-field, quintom-like gravitational
model. For the power-law model $f\left(  Q\right)  \simeq Q^{\frac{n}{n-1}}$,
we construct an analytic expression for the dynamical evolution of dark
energy, which depends on the parameter $n$. We constrain this dark energy
model using the the baryon acoustic oscillations from DESI DR2, and gamma-ray
bursts. The cosmological data provides $n\simeq0.33$. The $f\left(  Q\right)
$-model challenges the $\Lambda$CDM by providing a smaller value for
$\chi_{\min}^{2}$. Although the Akaike Information Criterion indicates that
the two cosmological models fit the data from the Pantheon+, the cosmic
chronometers and the baryonic acoustic oscillators in a similar way, when the
gamma-ray bursts are introduced there is weak evidence in favor of the
$f\left(  Q\right)  $-theory.

\end{abstract}
\keywords{Symmetric teleparallel gravity; $f\left(  Q\right)  $-theory; Dynamical dark
en ergy; Observational Constraints;}\maketitle

\section{Introduction}

\label{sec1}

The $f\left(  Q \right)  $-theory of gravity \cite{fq1,fq2}, along with other
extensions \cite{ex1,ex2,ex3,ex4} of Symmetric Teleparallel Equivalent to
General Relativity (STEGR) \cite{Nester:1998mp}, has been extensively studied
in recent years as a framework to describe cosmic acceleration and dark
energy. In STEGR, the gravitational field is described within a non-Riemannian
geometry. The connection is defined to be symmetric and flat, differing from
the Levi-Civita connection associated with the metric tensor. As a result, the
autoparallels are independent of the metric. The resulting gravitational field
equations in $f\left(  Q \right)  $-theory depend on both the nonlinear
function $f$ and the chosen connection, introducing new dynamical degrees of
freedom that govern the evolution and yield novel behaviors in the physical
parameters \cite{revh}.

In Friedmann--Lema\^{\i}tre--Robertson--Walker (FLRW) geometry, the definition
of the connection in $f(Q)$-gravity is not unique \cite{revh}. There exists a
branch of distinct gravitational models characterized by additional degrees of
freedom \cite{ndim1,av0}. These cosmological models are equivalent only in the
limit of the STEGR theory, that is, when the function $f\left(  Q \right)  $
is linear. Therefore, the definition of the connection is essential for
specifying the gravitational theory.

Most studies in spatially flat $f\left(  Q\right)  $-cosmology adopt the
connection defined in the coincidence gauge. This choice provides the minimum
number of dynamical degrees of freedom in the field equations
\cite{cf1,cf2,cf3,cf4,cf5,cf6,cf7,cf8,cf9,cf10,cf11,cf12,cf14,cf15,cf16,cf17}.
However, in this case, there exists a one-to-one correspondence between the
cosmological field equations and those of teleparallel $f\left(  T\right)
$-cosmology \cite{ft}, meaning that many known results from the literature are
simply recovered. This correspondence does not hold when non-trivial
connections are introduced. In such cases, new dynamical degrees of freedom
emerge, which can be attributed to scalar fields, unveiling new physics. The
scalar field description reveals that a quintom cosmological scenario can be
realized \cite{fqquintom}. Moreover, connections defined in the
non-coincidence gauge play a significant role in $f\left(  Q\right)  $-gravity
when nonzero spatial curvature is introduced \cite{vv0,vv1}, and in the study
of compact objects \cite{bh1,bh2,bh2b,bh3,bh4}.

In this work, we explore the viability of $f\left(  Q \right)  $-cosmology
with a non-trivial connection defined in the non-coincidence gauge, in the
context of cosmological observations. Specifically, we consider power-law
$f\left(  Q \right)  $-gravity and, by employing Hamilton-Jacobi theory, we
construct an analytic expression for the equation-of-state parameter of the
geometric dark energy component. Although $f\left(  Q \right)  $-theory is
challenged by issues such as strong coupling or the presence of ghosts in
cosmological perturbations \cite{ppr1,ppr2}, it serves as a useful case study
for understanding the role of the connection in gravitational dynamics. In
this sense, we treat it as a toy model to interpret cosmological data. We
consider the DESI DR2 BAO data \cite{des4,des5,des6} and gamma-ray bursts two
datasets that have been widely used in the analysis of dynamical dark energy
models
\cite{dd1,dd2,dd3,dd4,dd5,dd6,dd7,dd8,dd9,dd10,dd11,dd12,dd13,dd14,dd14a}. The
application of gamma-ray bursts in the problem of dark energy is widely
discussed in \cite{dd15}. The structure of the paper is as follows.

In Section \ref{sec2} we introduce the gravitational model of our
consideration, which is that of $f\left(  Q\right)  $-gravity in a spatially
flat FLRW geometry for a non-trivial connection. The resulting cosmological
model leads to the introduction of two scalar fields, which can be interpreted
as a phantom and a canonical scalar field within the minisuperspace framework.
For the power-law model $f\left(  Q\right)  \simeq Q^{\hat{n}}$,~$\hat{n}%
\neq0,1$, the cosmological field equations form an integrable Hamiltonian
system, and for a specific relation among the free parameters, we determine an
analytic expression for the geometric dynamical dark energy equation-of-state
parameter. This solution is used to construct an effective Hubble function. In
Section \ref{sec4}, we constrain the late-time cosmological observations using
the effective model and compare the results with those of the $\Lambda$CDM
model. Finally, in Section \ref{sec5}, we draw our conclusions.

\section{$f\left(  Q\right)  $-gravity}

\label{sec2}

We consider a spatially flat, isotropic, and homogeneous universe described by
the FLRW metric tensor%
\begin{equation}
ds^{2}=-N(t)^{2}dt^{2}+a(t)^{2}\left[  dr^{2}+r^{2}\left(  d\theta^{2}%
+\sin^{2}\theta d\varphi^{2}\right)  \right]  , \label{fq.14}%
\end{equation}
and the gravitational field is described by the symmetric and flat connection
\cite{mini1,Zhao}%
\[
\Gamma_{\;tt}^{t}=\frac{\ddot{\psi}(t)}{\dot{\psi}(t)}+\dot{\psi}(t),
\]%
\[
\Gamma_{tx}^{x}=\Gamma_{ty}^{y}=\Gamma_{tz}^{z}=\dot{\psi}\left(  t\right)  .
\]
The scalar field $\psi\left(  t\right)  $ describes the geometrodynamical
degrees of freedom related to the non-trivial connection.

Within the framework of $f\left(  Q\right)  $-gravity, for the metric tensor
(\ref{fq.14}) and the aforementioned connection, the cosmological field
equations describing the evolution of the physical parameters
are~\cite{f6,Hohmann,Heis2}%
\begin{gather}
3H^{2}f^{\prime}+\frac{1}{2}\left(  f(Q)-Qf^{\prime}\right)  +\frac{3\dot
{\psi}\dot{Q}f^{\prime\prime}}{2N^{2}}=0,\label{ff.01}\\
-\frac{2\left(  f^{\prime}H\right)  ^{\cdot}}{N}-3H^{2}f^{\prime}%
-\frac{\left(  f(Q)-Qf^{\prime}\right)  }{2}+\frac{3\dot{\psi}\dot{Q}%
f^{\prime\prime}}{2N^{2}}=0,\label{ff.02}\\
\dot{Q}^{2}f^{\prime\prime\prime}(+\left[  \ddot{Q}+\dot{Q}\left(
3NH-\frac{\dot{N}}{N}\right)  \right]  f^{\prime\prime}=0. \label{ff.03}%
\end{gather}
in which $H=\frac{1}{N}\frac{\dot{a}}{a}$ is the Hubble function and the
nonmetricity scalar $Q$ is defined as \cite{mini1,Zhao}%
\begin{equation}
Q=-6H^{2}+\frac{3\dot{\psi}}{N}\left(  3H-\frac{\dot{N}}{N^{2}}\right)
+\frac{3\ddot{\psi}}{N^{2}}. \label{ff.05}%
\end{equation}

Equations (\ref{ff.01}) and (\ref{ff.02}) are the modified Friedmann equations
of $f\left(  Q \right)  $-gravity, while equation (\ref{ff.03}) describes the
equation of motion for the scalar field $\psi\left(  t \right)  $. We remark
that for a linear $f\left(  Q \right)  $ function, i.e., $f\left(  Q \right)
= f_{1} Q + f_{0}$, the gravitational theory reduces to the limit of STEGR. On
the other hand, when $Q=Q_{0}$, then, the exact solution of STEGR is also recovered.

This set of cosmological field equations can be written in the equivalent form
of a quintom cosmological model. The scalar field $\psi\left(  t \right)  $
and the dynamical degrees of freedom related to the nonlinear function
$f\left(  Q \right)  $ correspond to one canonical and one phantom field in
the minisuperspace description \cite{fqquintom}.

\subsection{Scalar field description}

It has been found \cite{mini1} that the gravitational field equations
(\ref{ff.01}), (\ref{ff.02}), (\ref{ff.03}), and (\ref{ff.05}) admit a
minisuperspace description. Specifically, the dynamical degrees of freedom
related to $f\left(  Q\right)  $-gravity can be attributed to a scalar field.
Similar to the case of $f\left(  R\right)  $-gravity, we introduce the new
scalar field $\phi=f^{\prime}\left(  Q\right)  $ and the scalar field
potential $V\left(  \phi\right)  =\left(  f(Q)-Qf^{\prime}(Q)\right)  $. Then,
the gravitational field equations read
\begin{align}
3\phi H^{2}+\frac{3}{2N^{2}}\dot{\phi}\dot{\psi}+\frac{V\left(  \phi\right)
}{2}  &  =0,\label{ff.06}\\
-\frac{2}{N}\left(  \phi H\right)  ^{\cdot}-3\phi H^{2}+\frac{3}{2N^{2}}%
\dot{\phi}\dot{\psi}-\frac{V\left(  \phi\right)  }{2}  &  =0,\label{ff.07}\\
-3H^{2}+\frac{3}{2N}\left(  \dot{\psi}\left(  3H-\frac{\dot{N}}{N^{2}}\right)
+\frac{1}{N}\ddot{\psi}\right)  +\frac{V_{,\phi}}{2}  &  =0,\label{ff.08}\\
\frac{1}{N}\left(  \frac{1}{N}\dot{\phi}\right)  ^{\cdot}+\frac{3}{N}%
H\dot{\phi}  &  =0. \label{ff.09}%
\end{align}

The latter equations of motion follow from the variation of the point-like
Lagrangian%
\begin{equation}
L\left(  N,a,\dot{a},\phi,\dot{\phi},\dot{\psi}\right)  =-\frac{3}{N}\phi
a\dot{a}^{2}-\frac{3}{2N}a^{3}\dot{\phi}\dot{\psi}+\frac{N}{2}a^{3}V\left(
\phi\right)  . \label{lg2}%
\end{equation}

We observe that the field equations describe a constrained Hamiltonian system
with \thinspace$2\times3+1$ degrees of freedom. Without loss of generality, in
the following we assume $N = 1$, and expression (\ref{ff.06}) is interpreted
as the Hamiltonian constraint. \ 

Furthermore, equation (\ref{ff.09}) can be written in the form $\left(
a^{3}\dot{\phi}\right)  ^{\cdot}=0$, that is, the quantity $a^{3}\dot{\phi
}=I_{1}$ is conserved.

\subsection{Power-law theory}

Consider now the power-law potential $V\left(  \phi\right)  =-\frac{V_{0}}%
{6}\phi^{n},~n\neq0,1$, which corresponds to the power-law theory $f\left(
Q\right)  \simeq Q^{\hat{n}},~\hat{n}=\frac{n}{n-1}$.

For the latter function the cosmological field equations (\ref{ff.06}),
(\ref{ff.07}), (\ref{ff.08}) and (\ref{ff.09}) possess the additional
conservation law \cite{fqn2}%

\[
I_{2}=-4\left(  n+1\right)  \phi a^{2}\dot{a}+3\phi a^{3}\dot{\psi}.
\]

By using the latter conservation law and employing the Hamilton-Jacobi theory,
the field equations are reduced to the following system%
\begin{align}
\dot{a}  &  =-\frac{\left(  n+1\right)  I_{1}-K\left(  a,\phi\right)  }%
{6a^{2}\phi},\label{ll01}\\
\dot{\phi}  &  =\frac{I_{1}}{a^{3}},\label{ll02}\\
\dot{\psi}  &  =-\frac{2}{3}\frac{I_{1}\left(  n+1\right)  ^{2}-3I_{2}%
-\left\vert \left(  n+1\right)  \right\vert K\left(  a,\phi\right)  }%
{a^{3}\phi}.
\end{align}
where%
\[
K\left(  a,\phi\right)  =\sqrt{I_{1}^{2}\left(  n+1\right)  ^{2}-6I_{1}%
I_{2}+V_{0}a^{6}\phi^{n+1}}%
\]

We focus on the initial conditions with $6I_{1}I_{2}=I_{1}^{2}\left(
n+1\right)  ^{2}$. We determine the parametric solution%
\begin{equation}
\phi\left(  a\right)  =\left(  -\frac{\sqrt{V_{0}}}{I_{1}\left(  n+1\right)
}a^{3}W\left(  I_{0}a^{-3}\right)  \right)  ^{-\frac{2}{1+n}}, \label{ll.04}%
\end{equation}
where $W\left(  x\right)  $ is the Lambert function and $I_{0}$ is an
integration constant.

Thus, the corresponding equation of state parameter $w_{f\left(  Q\right)
}\left(  a\right)  =-1-\frac{2}{3}\frac{a}{H}\frac{dH}{da}$ is calculated
\begin{equation}
w_{f\left(  Q\right)  }\left(  a\right)  =-1+\frac{2W\left(  I_{0}%
a^{-3}\right)  \left(  2n+\left(  n-1\right)  W\left(  I_{0}a^{-3}\right)
\right)  }{\left(  n+1\right)  W\left(  I_{0}a^{-3}\right)  ^{2}}.
\label{ll.05}%
\end{equation}

\begin{figure*}[ptbh]
\centering\includegraphics[width=0.7\textwidth]{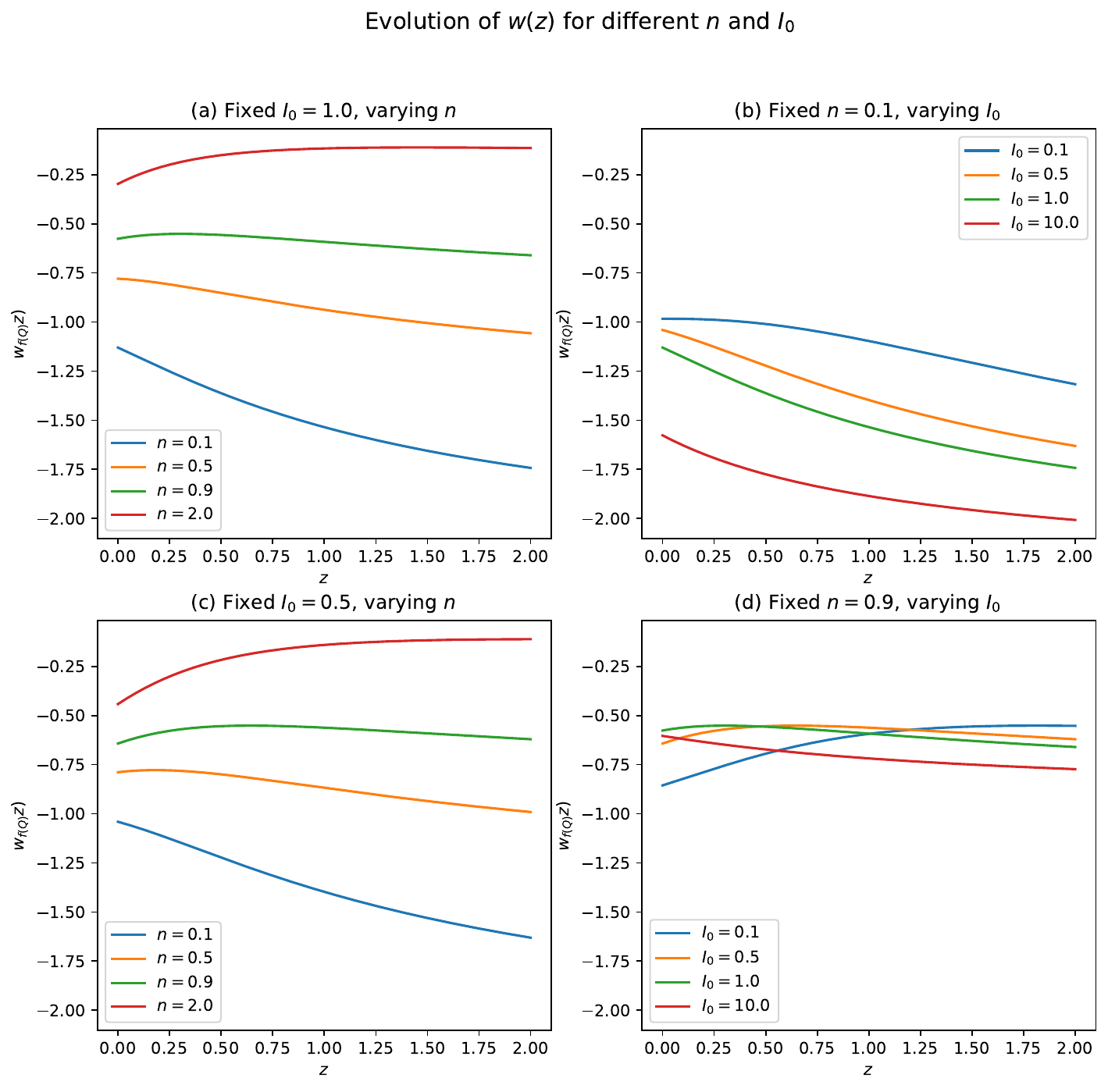}\caption{Qualitative
evolution of the equation of state parameter$~w_{f\left(  Q\right)  }\left(
z\right)  $ given by expression (\ref{ll.05}) for different values of the free
parameters $n$ and $I_{0}$. }%
\label{fig1}%
\end{figure*}

\begin{figure*}[ptbh]
\centering\includegraphics[width=0.7\textwidth]{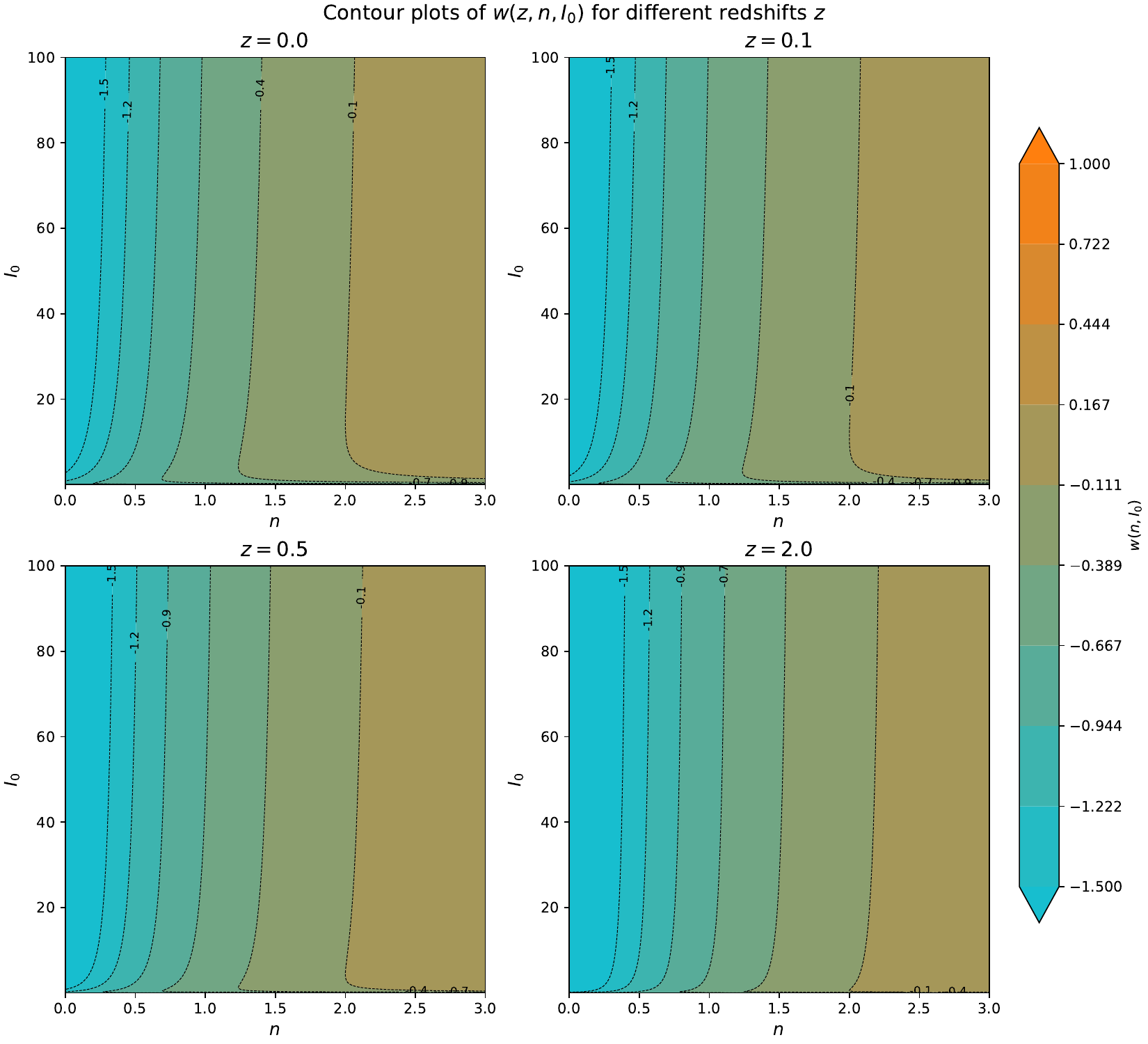}\caption{Contour
plots on the parametric space $\left(  n,I_{0}\right)  $ for the equation of
state parameter$~w_{f\left(  Q\right)  }\left(  z\right)  $ given by
expression (\ref{ll.05}) for different for specific redshifts $z$.}%
\label{fig2}%
\end{figure*}

The equation-of-state parameter depends on two constants: the power $n$ and
the integration constant $I_{0}$. In order to understand how the
equation-of-state parameter $w_{f\left(  Q\right)  }$ behaves for different
values of the free parameters, in Fig. \ref{fig1} we present the qualitative
evolution of the equation-of-state parameter (\ref{ll.05}) for various values
of $n$ and $I_{0}$. Moreover, in Fig. \ref{fig2}, we provide contour plots in
the two-dimensional space $\left(  n,I_{0}\right)  $ for specific values of
the redshift.

As expected, the main evolution of $w_{f\left(  Q \right)  }$ is independent
of the value of $I_{0}$, especially for large values $I_{0} \gg1$, where
$w_{f\left(  Q \right)  }$ becomes almost independent. This can be seen by
performing the change of variable $I_{0} a^{-3} \rightarrow\alpha$. Thus, in
the following, we consider $I_{0} = 1$.

Furthermore, for the plots it follows that for $0<n<1$, i.e. $\hat{n}<0,$
$w_{f\left(  Q\right)  }$ cross phantom divide line, as it was found before in
\cite{fqdyn}.

It is important to mention that this is an analytic solution for the power-law
case; however, it can also be viewed as an asymptotic solution for any
function $f\left(  Q \right)  $ in which the power-law term dominates. We make
use of this property in the present work to define an analytic expression for
the Hubble function.

Indeed, in the presence of a dust fluid source which attributes the dark
matter the Lagrangian function (\ref{lg2}) is modified
\begin{equation}
\hat{L}=-\frac{3}{N}\phi a\dot{a}^{2}-\frac{3}{2N}a^{3}\dot{\phi}\dot{\psi
}+\frac{N}{2}a^{3}V\left(  \phi\right)  -N\rho_{m0}\phi a^{-3}.
\end{equation}
The coupling of the dust fluid with the scalar field $\phi$, has been
introduced necessary in order the Friedmann equations (\ref{ff.01}),
(\ref{ff.02}) to read%
\begin{gather}
3H^{2}=\rho_{m0}a^{-3}+\rho_{f\left(  Q\right)  },\\
-2\dot{H}-3H^{2}=p_{fQ},
\end{gather}
where $\rho_{m0}$ is the energy density for the dust fluid, and $\rho
_{f\left(  Q\right)  }$, $p_{f\left(  Q\right)  }$ are the effective
\begin{align}
\rho_{f\left(  Q\right)  }  &  =-\left(  \frac{3}{2N^{2}}\frac{\dot{\phi}%
}{\phi}\dot{\psi}+\frac{1}{2\phi}V\left(  \phi\right)  \right) \\
p_{f\left(  Q\right)  }  &  =-2\frac{\dot{\phi}}{\phi}H-\frac{3}{2N^{2}}%
\frac{\dot{\phi}}{\phi}\dot{\psi}+\frac{1}{2}\frac{V\left(  \phi\right)
}{\phi}.
\end{align}
Hence, the asymptotic solution for the Hubble function for the power-law
function $f\left(  Q\right)  \simeq Q^{\hat{n}}$ reads
\begin{equation}
\frac{H^{2}\left(  a\right)  }{H_{0}^{2}}=\left(  \Omega_{m0}a^{-3}+\left(
1-\Omega_{m0}\right)  e^{\left(  -3\int_{a_{0}}^{a}\frac{1+w\left(
\alpha\right)  }{\alpha}d\alpha\right)  }\right)  . \label{ll.06}%
\end{equation}
where $w\left(  a\right)  $ is given from expression (\ref{ll.05}). At this
point we not that the latter Hubble function is an exact solution when
$\Omega_{m0}\rightarrow0$.

In the following Section we make us of cosmological observations to constraint
the latter Hubble function and determine the best-fit values for the free
parameters, $H_{0}$,~$\Omega_{m0}$ and the power $n$.

\section{Observational Constraints}

\label{sec4}

The dataset that we employ in this study are

\begin{itemize}
\item Observational Hubble Data (OHD): These data include direct measurements
of the Hubble parameter, without any cosmological assumptions, obtained from
the differential age evolution of cosmic chronometers. Cosmic chronometers are
old, passively evolving galaxies with synchronous stellar populations and
similar cosmic evolution \cite{co01}. In this work, we use 31 direct
measurements of the Hubble parameter for redshifts in the range $0.09 \leq z
\leq1.965$~\cite{cc1}.

\item Pantheon+ Supernova: This dataset includes 1701 light curves of 1550
spectroscopically confirmed supernova events within the range $10^{-3}%
<z<2.27~$\cite{pan}. We consider the Pantheon+ data with the Supernova $H_{0}$
for the Equation of State of Dark energy Cepheid host distances calibration
(SN$_{0}$) and without the Cepheid calibration (SN). The luminosity
distance~$D_{L}=c\left(  1+z\right)  \int\frac{dz}{H\left(  z\right)  }$, is
used to define the theoretical distance modulus$~\mu^{th}=5\log D_{L}+25$,
which is used to constraint with the distance modulus $\mu^{obs}~$at~observed
redshifts~$z$.

\item Gamma-ray bursts (GRB): \ We consider the 193 events where they have
with the use of the Amati correlation \cite{amm} in the redshift range
$0.0335<z<8.1$. The distance modulus $\mu^{obs}$ for each event at the
observed redshift are summarized in \cite{amti}.

\item Baryonic acoustic oscillations (BAO): These data are data from the DESI
DR2 release \cite{des4,des5,des6}. The dataset includes observation values of
the $\frac{D_{M}}{r_{d}}=\frac{\left(  1+z\right)  ^{-1}D_{L}}{r_{d}}%
,~\frac{D_{V}}{r_{d}}=\frac{\left(  cD_{L}\frac{z}{H\left(  z\right)
}\right)  ^{\frac{1}{3}}}{r_{d}}$~and$~\frac{D_{H}}{r_{d}}=\frac{c}{r_{d}H}$
where $r_{d}$ is the sound horizon at the drag epoch, $D_{M}$ is the comoving
angular distance; $D_{V}$ is the volume averaged distance and $D_{H}$ is the
Hubble distance.
\end{itemize}

We use the Bayesian inference COBAYA\footnote{https://cobaya.readthedocs.io/}
\cite{cob1,cob2} with a custom theory and the MCMC sampler. We constraint the
Hubble function (\ref{ll.06}) for the free parameters $\left\{  H_{0}%
,\Omega_{m0},n,r_{d}\right\}  $. Furthermore, we compare our findings with the
$\Lambda$CDM which has the free parameters $\left\{  H_{0},\Omega_{m0}%
,r_{d}\right\}  $.

Since the parameter spaces of the two models differ in dimensionality, the
Akaike Information Criterion (AIC)~\cite{AIC} and the Bayesian Information
Criterion (BIC) are used to enable a comparison.

The AIC for Gaussian errors and for large number of data is defined as%
\[
AIC\simeq-2\ln\mathcal{L}_{\max}+2\kappa,
\]
where~$\mathcal{L}_{\max}$ is the maximum value for the likelihood
$\mathcal{L}_{\max}=\exp\left(  -\frac{1}{2}\chi_{\min}^{2}\right)  $. For
combined data, it follows
\[
\mathcal{L}_{\max}^{total}=\mathcal{L}_{\max}^{SN}\times\mathcal{L}_{\max
}^{CC}\times\mathcal{L}_{\max}^{BAO}.
\]

From the difference of the AIC parameters for two models%
\[
\Delta\left(  AIC\right)  =AIC_{1}-AIC_{2}%
\]
it can be concluded whether the models are statistically distinguishable. A
larger absolute difference in AIC indicates stronger evidence in favor of the
model with the lower AIC value. Specifically, a difference $\left\vert
\Delta\left(  AIC\right)  \right\vert >2$ indicates positive evidence, while
$\left\vert \Delta\left(  AIC\right)  \right\vert >6$ indicates strong
evidence. On the other hand, a value $\left\vert \Delta\left(  AIC\right)
\right\vert <2$ suggests that the two models are statistically consistent with
each other.

Furthermore, the BIC is defined as
\[
BIC\simeq-2\ln\mathcal{L}_{\max}+\kappa\ln N,
\]
where $N$ is the dimension of the data size. The difference of the BIC
parameters for two models is
\[
\Delta\left(  BIC\right)  =BIC_{1}-BIC_{2}.
\]
A difference of$~\left\vert \Delta\left(  BIC\right)  \right\vert >2$
indicates positive evidence, while $\left\vert \Delta\left(  BIC\right)
\right\vert >6$ indicates strong evidence in favor of the model with lower
value of BIC. Finally, a value $\left\vert \Delta\left(  BIC\right)
\right\vert <2$ suggests that the two models are statistically consistent with
each other.

\subsection{Results}

We constrain the cosmological theory (\ref{ll.06}) using different
combinations of datasets. The best-fit parameters with the corresponding
$1\sigma$ uncertainties, along with the comparison to the $\Lambda$CDM model,
are presented in Table \ref{data1}.

For the dataset SN+OHD+BAO, we find the best-fit parameters: $H_{0}%
=68.0_{-1.7}^{+1.7}$, $\Omega_{m0}=0.315_{-0.012}^{+0.013}$, $n=0.328_{-0.05}%
^{+0.05}$ and $r_{d}=147.2_{-3.6}^{+3.6}$. The differences in the minimum
chi-squared and AIC values are $\chi_{\min}^{2}-\chi_{\Lambda\min}^{2}=-2.1$,
$AIC-AIC_{\Lambda}=-0.1$,~and $BIC-BIC_{\Lambda}=+5.4$ respectively. Although
$f\left(  Q\right)  $-gravity yields a smaller value of $\chi_{\min}^{2}$, due
to the difference in degrees of freedom, both models fit the data equally
well. However, from the BIC supports the $\Lambda$CDM.

However, when the GRB dataset is included, the best-fit parameters remain
within $1\sigma$ are $H_{0}=68.4_{-1.7}^{+1.7}$, $\Omega_{m0}=0.302_{-0.013}%
^{+0.013}$, $n=0.334_{-0.047}^{+0.047}~$and $r_{d}=147.6_{-3.5}^{+3.5}$. In
this case, we find $\chi_{\min}^{2}-\chi_{\Lambda\min}^{2}=-3.9$,
$AIC-AIC_{\Lambda}=-1.9$ and $BIC-BIC_{\Lambda}=+3.7$, indicating that
$f\left(  Q\right)  $-gravity fits the GRB data better than $\Lambda$CDM.
According to the AIC, there is weak evidence in favor of the power-law
$f\left(  Q\right)  $-gravity over $\Lambda$CDM, however the BIC indicate a
positive evidence in favor of the simplest model, that is of the $\Lambda$CDM.

When we consider the Cepheid calibration for the supernovae, the best-fit
parameters for the dataset SN$_{0}$+OHD+BAO are $H_{0}=72.04_{-0.89}^{+0.89}$,
$\Omega_{m0}=0.311_{-0.012}^{+0.012}$, $n=0.308_{-0.046}^{+0.046}$ and
$r_{d}=139.8_{-2.0}^{+2.0}$. The corresponding statistical comparisons yield
$\chi_{\min}^{2}-\chi_{\Lambda\min}^{2}=-1.9$,~ $AIC-AIC_{\Lambda}=+0.2$ and
$BIC-BIC_{\Lambda}=+7.5$. As before, the inclusion of GRB data supports the
power-law $f\left(  Q\right)  $-gravity. The best-fit parameters in this case
are $H_{0}=72.15_{-0.88}^{+0.88}$, $\Omega_{m0}=0.298_{-0.012}^{+0.012}$,
$n=0.318_{-0.045}^{+0.045}~$and $r_{d}=140.7_{-2.0}^{+1.8}$, with $\chi_{\min
}^{2}-\chi_{\Lambda\min}^{2}=-3.3$, $AIC-AIC_{\Lambda}=-1.3$ and
$BIC-BIC_{\Lambda}=+4.3$. The AIC states that the two models fit the data in
similar way, while the BIC suggests a positive evidence in favor of the
$\Lambda$CDM.

\begin{figure}[ptbh]
\centering\includegraphics[width=0.5\textwidth]{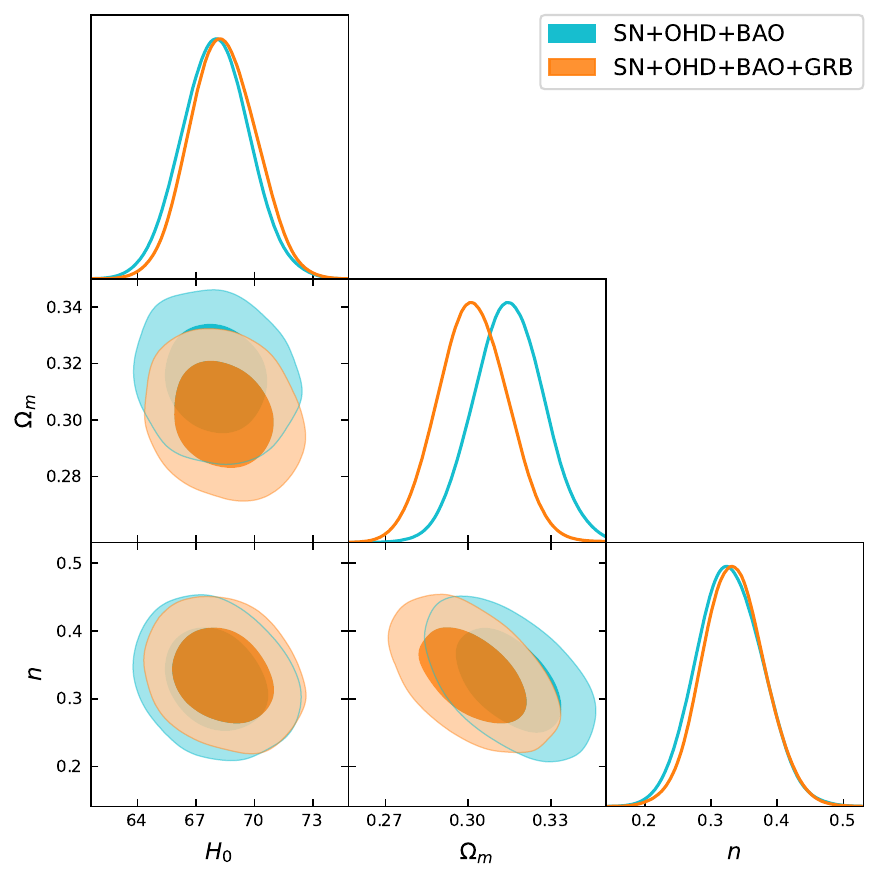}\caption{Confidence
space for the best-fit parameters for the power-law $f\left(  Q\right)
$-gravity, for the datasets SN+OHD+BAO and SN+OHD+BAO+GRB.}%
\label{fig001}%
\end{figure}

\begin{figure}[ptbh]
\centering\includegraphics[width=0.5\textwidth]{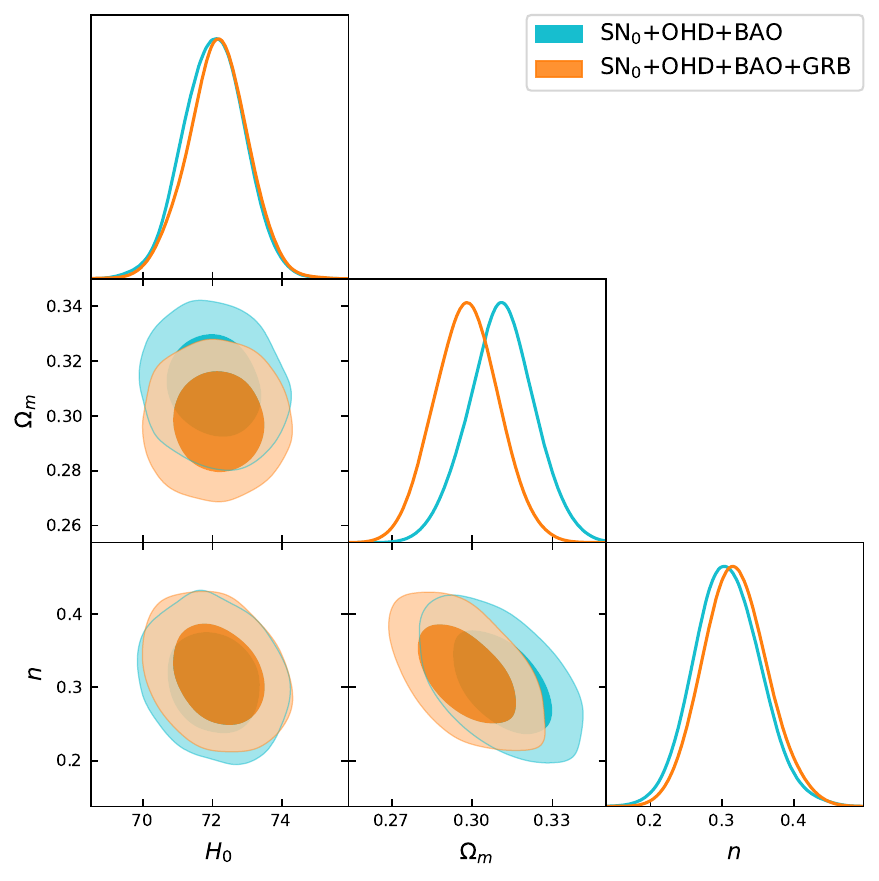}\caption{Confidence
space for the best-fit parameters for the power-law $f\left(  Q\right)
$-gravity, for the datasets SN$_{0}$+OHD+BAO and SN$_{0}$+OHD+BAO+GRB with the
Cepheid calibration for the supernova.}%
\label{fig002}%
\end{figure}

In Figs. \ref{fig001} and \ref{fig002} we present the confidence space for the
best-fit parameters for the power-law $f\left(  Q\right)  $-model.

Although in this work we have considered the value of the integration constant
$I_{0}=1,~$in (\ref{ll.05}), we performed the same constraints and for other
values of $I_{0}$. The physical parameters and the $\chi_{\min}^{2}$ remain
unchanged, while there is a small change in the best-fit value for parameter
$n$, as expected from the contour plots of Fig. \ref{fig2}.%

%TCIMACRO{\TeXButton{B}{\begin{table}[tbp] \centering}}%
%BeginExpansion
\begin{table*}[tbp] \centering
%EndExpansion
\caption{Best fit parameters of the power-law model.}%
\begin{tabular}
[c]{cccccccc}\hline\hline
$\mathbf{f}\left(  Q\right)  \simeq Q^{\frac{n}{n-1}}$ & $\mathbf{H}${$_{0}$}
& $\mathbf{\Omega}_{m0}$ & $\mathbf{n}$ & $\mathbf{r}_{d}$ & $\mathbf{\chi
}_{\min}^{2}-\mathbf{\chi}_{\Lambda\min}^{2}$ & $\mathbf{AIC}-\mathbf{AIC}%
_{\Lambda}$ & $\mathbf{BIC}-\mathbf{BIC}_{\Lambda}$\\
\textbf{SN+OHD+BAO} & $68.0_{-1.7}^{+1.7}$ & $0.315_{-0.012}^{+0.013}$ &
$0.328_{-0.05}^{+0.05}$ & $147.2_{-3.6}^{+3.6}$ & $-2.1$ & \thinspace$-0.1$ &
$+5.4$\\
\textbf{SN+OHD+BAO+GRB} & $68.4_{-1.7}^{+1.7}$ & $0.302_{-0.013}^{+0.013}$ &
$0.334_{-0.047}^{+0.047}$ & $147.6_{-3.5}^{+3.5}$ & $-3.9$ & $-1.9$ & $+3.7$\\
\textbf{SN}$_{0}$\textbf{+OHD+BAO} & $72.04_{-0.89}^{+0.89}$ & $0.311_{-0.012}%
^{+0.012}$ & $0.308_{-0.046}^{+0.046}$ & $139.8_{-2.0}^{+2.0}$ & $-1.8$ &
$+0.2$ & $+7.5$\\
\textbf{SN}$_{0}$\textbf{+OHD+BAO+GRB} & $72.15_{-0.88}^{+0.88}$ &
$0.298_{-0.012}^{+0.012}$ & $0.318_{-0.045}^{+0.045}$ & $140.7_{-2.0}^{+1.8}$
& $-3.3$ & $-1.3$ & $+4.3$\\\hline\hline
\end{tabular}
\label{data1}%
%TCIMACRO{\TeXButton{E}{\end{table}}}%
%BeginExpansion
\end{table*}%
%EndExpansion

\section{Conclusions}

\label{sec5}

In this study, we examined the cosmological viability of the non-coincidence
$f\left(  Q \right)  $-theory using late-time cosmological observations.
Specifically, we considered an FLRW universe with a non-trivial connection,
where the effective dark energy component in $f\left(  Q \right)  $-theory
exhibits quintom behavior, that is, it is described by two scalar fields: one
canonical and one phantom. For the power-law model $f\left(  Q \right)  \simeq
Q^{\frac{n}{n-1}}$, the field equations are integrable, and we were able to
derive a closed-form solution for the effective dark energy equation-of-state parameter.

We employed the Pantheon+ supernova dataset, the direct measurements of the
Hubble parameter from cosmic chronometers, the baryonic acoustic oscillations
from DESI DR2, and the gamma-ray bursts. The Bayesian analysis of these data
indicates that the parameter $n\simeq0.334_{-0.047}^{+0.047}$, leading to a
negative index in the $f\left(  Q\right)  $ model. This model should not be
confused with the one based on the coincidence gauge. Our model and the
coincidence gauge model are not dynamically or physically equivalent.

Comparing the best-fit parameters with those of $\Lambda$CDM, we conclude that
the $f\left(  Q \right)  $-theory provides a better fit to the cosmological
data, while the AIC indicates weak evidence in favor of the $f\left(  Q
\right)  $-theory. This evidence becomes stronger when gamma-ray bursts are
included in the analysis.

Until now, $f\left(  Q \right)  $-theory has primarily been tested against
cosmological observations in the coincidence gauge, often reproducing results
already known from $f\left(  T \right)  $-gravity. To the best of our
knowledge, this is the first analysis of its kind involving an exact
cosmological solution of $f\left(  Q \right)  $-gravity with a non-coincidence
connection. The results of this work suggest that the non-coincidence
connection in $f\left(  Q \right)  $-theory may play a significant role in the
description of dark energy.

\begin{acknowledgments}
The author thanks the support of VRIDT through Resoluci\'{o}n VRIDT No.
096/2022 and Resoluci\'{o}n VRIDT No. 098/2022. Part of this study was
supported by FONDECYT 1240514 ETAPA 2025.
\end{acknowledgments}

\end{document}